\documentclass[10pt,aps,prd,showpacs,twocolumn,superscriptaddress,nofootinbib,floatfix]{revtex4-1}
\usepackage{amsfonts,amsmath,amssymb,amsthm}
\usepackage[bookmarks, bookmarksopen, bookmarksnumbered]{hyperref}
\usepackage{graphicx}
\usepackage{tensor}
\usepackage{color}
\usepackage{epigraph}
\begin{document}

\title{Stationary massive disks around black holes: realistic equation of state and bifurcation}

\author{Wojciech Kulczycki}
\affiliation{Instytut Fizyki Teoretycznej, Uniwersytet Jagiello\'nski, {\L}ojasiewicza 11, 30-348 Krak\'{o}w, Poland}

\author{Patryk Mach}
\affiliation{Instytut Fizyki Teoretycznej, Uniwersytet Jagiello\'nski, {\L}ojasiewicza 11, 30-348 Krak\'{o}w, Poland}

\author{Edward Malec}
\affiliation{Instytut Fizyki Teoretycznej, Uniwersytet Jagiello\'nski, {\L}ojasiewicza 11, 30-348 Krak\'{o}w, Poland}

\begin{abstract}
We   study stationary and axially symmetric  black hole-disk systems, assuming a combination of  the DD2 and Timmes-Swesty equations of state and a three-parameter family of rotation laws. There exist two branches of solutions that are shown to bifurcate, for a suitable specific entropy   and a parameter in the rotation law. Low entropy nuclear matter allows for the existence of moderately massive Keplerian disks.
\end{abstract}

\maketitle
 
\section{Introduction}

We shall investigate in this paper a stationary axially symmetric system consisting of a black hole and a toroid. The matter in the toroid is a perfect fluid that satisfies a tabulated equation of state of \cite{sho_winter}.

The description of such systems can be self-contained, provided that a rotation law is supplied. The main challenge is to find a rotation law that is realistic and at the same time  solvable numerically. Uniformly (rigidly) rotating gaseous disks in general-relativistic hydrodynamics have  been  discussed in \cite{Bardeen,Ipser}. A more realistic angular velocity profile has been  studied since 1980's---in the context of rotating stars---with the angular momentum density being  a linear function of the frequency  \cite{komatsu,nishida_eriguchi,nishida1}. We should mention important investigations of the rigid rotation by the Jena group \cite{NM,AP,MAKNP}. Later various nonlinear differential rotation laws have been proposed and implemented numerically \cite{GYE,UTG,UTB,TUS}. A rotation law describing disks in motion around   black holes has been obtained in \cite{MM}. In what follows we focus on its generalization---a new family of rotation laws that was derived recently \cite{KM2020}. This family includes  the Keplerian rotation, by which  we mean a special case of the rotation law derived in \cite{kkmmop}---see Eq.\ (\ref{keplerian_rl}) below. 

There are two reasons to study toroids circulating around black holes. Firstly, the numerical simulations of the coalescence of two neutron stars indicate, that there might appear a quasi-stationary phase with remnants consisting of a black hole surrounded by a toroid. We recall reviews of different scenarios that can lead to that picture   \cite{shibata2019,stergioulas2020}. This  system should be further evolved in order to make predictions concerning related electromagnetic or gravitational-wave observation. These coalescence simulations are numerically expensive, and it would be advantageous to start from initial data supplied by a (suitably deformed and supplied with a relevant physics) stationary black hole-toroid configuration. Such a programme has already been implemented, see recent results in \cite{sho_winter, sho_autumn}. Thus there is a need for a catalogue of stationary solutions that could serve as idealizations of the   remnants that are produced in real merger processes. Secondly, the mathematics of black hole-disk configurations is interesting. They are described by a free-boundary system of nonlinear elliptic equations, and one can expect that there shall appear typical phenomena such as nonuniqueness of solutions or bifurcations, which in turn can be associated with the emergence of instability. 

The order of the main part of the paper is as follows.
The second  Section is dedicated to the description of equations. Section \ref{numerics} explains the tabulated equation of state within the toroidal matter.
Section \ref{rotation} explains the relation between our family of laws and those of Fujibayashi et al.\ \cite{sho_winter}. Next Section is dedicated to the description of numerics. In particular,
we test in Sec.\ \ref{sec:Vtest} our numerical procedure, using the rotation law of  \cite{sho_winter}. We get a satisfactory agreement with results of \cite{sho_winter}  in all analysed cases.  We compare also disk  solutions that satisfy the same boundary data, but different rotation laws---those of \cite{sho_winter} and of \cite{KM2020}.  We describe examples of bifurcation in Sec.\ \ref{bifurcation}, with the bifurcation parameter being the specific entropy $s$ or the exponent parameter $\delta $ of the rotation law, respectively. Section \ref{Kepler} is dedicated mainly to the discussion of Keplerian rotation laws. It is known that the rotation law of \cite{sho_winter} excludes compact Keplerian solutions with light disks. In contrast to that, a Keplerian law of \cite{KM2020} does allow for compact  solutions with light disks for a range of the specific entropy. These tori can gain a mass of the   order of $0.06 M_\odot $ if the specific entropy $s$ is relatively low.

\section{Stationary toroids around black holes: equations  }
 
The formulation and numerical  methods of  this paper are based on  a scheme of \cite{shibata}. They have been used (with some changes) also in \cite{kkmmop,kkmmop2,kmm,dyba}. In this section we only give a brief description of the key elements of the formalism.

We use (with a few exceptions) standard gravitational system of units with $c = G = 1$, where $c$ is the speed of light, and $G$ is the gravitational constant. The signature of the metric is   $(-,+,+,+)$.   Greek indices are used to label spacetime dimensions, $\mu , \nu, \ldots = 0,1,2,3$. Spatial dimensions are denoted by Latin indices $i, j, \ldots = 1,2,3$.

\subsection{Hydrodynamical equations}

We shall use stationary, axially symmetric metrics of the form
\begin{equation}
\label{generalmetric}
g = g_{tt} dt^2 + 2 g_{t \varphi} dt d\varphi + g_{rr} dr^2 + g_{\theta \theta} d\theta^2 + g_{\varphi \varphi} d\varphi^2, 
\end{equation}
where $(t,r,\theta,\varphi)$ denote spherical coordinates, and where the components of the metric tensor $g_{tt}$, $g_{t\varphi}$, $g_{rr}$, $g_{\theta \theta}$, $g_{\varphi \varphi}$ depend only on $r$ and $\theta$.
Because of numerical convenience, in the majority of this work we will specialize to the following quasi-isotropic gauge
\begin{eqnarray}
\nonumber
g & = & - \alpha^2 dt^2 + \psi^4 e^{2q} (dr^2 + r^2 d\theta^2) + \\
&& \psi^4 r^2 \sin^2 \theta (\beta dt + d \varphi)^2.
\label{isotropic}
\end{eqnarray}
There exist two independent Killing vectors, azimuthal (rotational) and asymptotically timelike, with contravariant components $\eta^\mu = (0,0,0,1)$ and $\xi^\mu = (1,0,0,0)$, respectively.

We assume the energy-momentum tensor of the perfect fluid
\begin{equation}
T^{\mu \nu} = \rho h u^\mu u^\nu + p g^{\mu \nu},
\end{equation}
where $\rho$ is the rest-mass density, $h$ is the specific enthalpy, and $p$ is the pressure. The four-velocity of the fluid $u^\mu$ is normalized: $u_\mu u^\mu = -1$.  

In what follows we shall deal only with the azimuthal stationary rotation:   $u^\mu = (u^t,0,0,u^\varphi) = u^t(1,0,0,\Omega)$. The component $u^t$ can be expressed in terms of the angular velocity $\Omega = u^\varphi/u^t$ as
\begin{equation}
\label{ut}
(u^t)^2 = - \frac{1}{g_{tt} + 2 g_{t \varphi} \Omega + g_{\varphi \varphi} \Omega^2}.
\end{equation}
The assumptions of stationarity and axial symmetry imply that $u^t$, $u^\varphi$, $\rho$, $p$, and $h$ can only depend on $r$ and $\theta$.

For a barotropic fluid, the conservation equations
\begin{equation}
\label{cons_eqs_general}
\nabla_\mu \left( \rho u^\mu \right) = 0, \quad \nabla_\mu T^{\mu \nu} = 0
\end{equation}
can be integrated, assuming that the angular momentum per unit inertial mass, $j = u^t u_\varphi$, depends only on the angular velocity $\Omega$. In this case, one obtains in the region of nonnegative mass density ($\rho > 0$)
\begin{equation}
\label{bernoulli3}
\ln \left( \frac{h}{u^t} \right) + \int j(\Omega) d\Omega  = C,
\end{equation}
where $C$ denotes an integration constant. We will refer to Eq.\ (\ref{bernoulli3}) as the Euler-Bernoulli equation.

We should warn the reader, that   an alternative choice is to define the angular momentum density $\tilde j=hu_\varphi $; that also leads to the  equation analogous to (\ref{bernoulli3})---of the following form: 

\begin{equation}
\label{bernoulli4}
\frac{h}{u^t} + \int \tilde  j(\Omega) d\Omega  = C_1.
\end{equation}
This option is adopted in \cite{sho_winter} and \cite{sho_autumn}.

In the main part of this paper we assume  the recently derived rotation law \cite{KM2020} 

 \begin{eqnarray}
\label{momentum}
j(\Omega ) &  \equiv &  -\frac{1}{\kappa (1+\delta)}\frac{d}{d\Omega} \ln \left[  1-  (a_\mathrm{rot}\Omega)^2\right. \nonumber\\
&&   \left.- \kappa  w^{1- \delta }\Omega^{1+\delta }(1-a_\mathrm{rot}\Omega)^{1-\delta} \right],
\end{eqnarray}
 where $w$ is a   constant, and $\delta$, $\kappa$ and $a_\mathrm{rot}$ are parameters.  The value of $w$ is obtained in the process of solving the relevant equations, as explained  later. Note that in the Newtonian limit for $\kappa=(1-3\delta)/(1+\delta)$, Eq.\ (\ref{momentum}) yields $\Omega = w/(r \sin \theta)^\frac{2}{1-\delta}$.

 The special case of this formula is the Keplerian rotation law that corresponds to the choice of $\delta=-1/3$ and $\kappa=(1-3\delta)/(1+\delta)$, i.e.,
\begin{eqnarray}
j(\Omega) & = &  \frac{a_\mathrm{rot}^2 \Omega^\frac{4}{3} + w^\frac{4}{3} (1 - 3 a_\mathrm{rot} \Omega) (1 - a_\mathrm{rot} \Omega)^\frac{1}{3}}{\Omega^\frac{1}{3} \left[ 1 - a_\mathrm{rot}^2 \Omega^2 - 3 w^\frac{4}{3} \Omega^\frac{2}{3}
   (1-a_\mathrm{rot} \Omega)^\frac{4}{3} \right]} \nonumber \\
& = & - \frac{1}{2} \frac{d}{d \Omega} \ln \left\{ 1 - \left[ a_\mathrm{rot}^2 \Omega^2 + 3 w^\frac{4}{3} \Omega^\frac{2}{3} (1 - a_\mathrm{rot} \Omega)^\frac{4}{3} \right] \right\}. \nonumber \\&&
\label{keplerian_rl}
\end{eqnarray}
This formula was obtained in \cite{kkmmop, kkmmop2}, where its physical relevance was thoroughly discussed. It has been recently applied in \cite{kmm, mgfop, dyba}. Massless disks of dust around a Kerr black hole are subjected to  the Keplerian rotation with the parameter $a_\mathrm{rot}=a$, where $a$ is the black hole spin parameter (see Sec.\ \ref{einstein_equations} for the definition). 
 
In former calculations we often had chosen $a_\mathrm{rot}$ to be equal to the spin parameter of the black hole $a$.
Hereafter we decided to consider also the case $a_\mathrm{rot}\ne a$, for a reason that is to be explained later, in Section \ \ref{Kepler}.  The circular geodesic motion of a test body in the equatorial plane of the Kerr spacetime with the mass $m$ and spin $a$ is given by Eq.\ (\ref{keplerian_rl}) with $w^2 = m$ and $a_\mathrm{rot}=a$. In the case of self-gravitating toroids, however, $w^2 \neq m$. Equation (\ref{keplerian_rl}) gives the Keplerian angular velocity $\Omega = w/(r \sin \theta)^\frac{3}{2}$ in the Newtonian limit.

In the main part of this work we will use the rotation law (\ref{momentum}) with free parameters $\delta$ and $a_\mathrm{rot}$, and $\kappa = (1-3\delta)/(1+\delta)$, as motivated by the Keplerian rotation law. This rotation law will be further referred to as $j_\mathrm{KM}$:
\begin{eqnarray}
j_\mathrm{KM}(\Omega ) &  \equiv &  -\frac{1}{(1 - 3\delta)}\frac{d}{d\Omega} \ln \left[  1-  (a_\mathrm{rot}\Omega)^2\right. \nonumber\\
&&   \left.- \frac{1-3\delta}{1+\delta}  w^{1- \delta }\Omega^{1+\delta }(1-a_\mathrm{rot}\Omega)^{1-\delta} \right].
\label{kmm}
\end{eqnarray}

Given the relation $j(\Omega)$ and the metric, one can compute the angular velocity $\Omega$ by solving the equation
\begin{equation}
\label{rot_law_eq}
j(\Omega) \left[ \alpha^2 - \psi^4 r^2 \sin^2 \theta (\Omega + \beta)^2 \right] = \psi^4 r^2 \sin^2 \theta (\Omega + \beta),
\end{equation}
which is directly implied  by the definition $j = u^t u_\varphi$. In the following, we assume a convention with $\Omega > 0$. The torus would be said to  corotate, if $a>0$, and counterrotate, for $a < 0$.

Taking the above definitions into account, one can write the Euler-Bernoulli Eq.\ (\ref{bernoulli3}) as
\begin{eqnarray}
\label{bernoulli2}
&& C^\prime = \lefteqn{h \sqrt{ \alpha^2 - \psi^4 r^2 \sin^2 \theta (\Omega + \beta)^2}} \\
&&  \times \left\{ 1 - \left[ a_\mathrm{rot}^2 \Omega^2 + \kappa w^{1-\delta} \Omega^{1+\delta} (1 - a_\mathrm{rot} \Omega)^{1-\delta} \right] \right\}^{-\frac{1}{\kappa(1+\delta)}},\nonumber
\end{eqnarray}
where $C^\prime$ is a constant.

\subsection{Einstein equations}
\label{einstein_equations}

The spacetime metric is not given by the Kerr solution for self-gravitating tori, but the Kerr metric plays an important role in our construction. We shall need the Kerr metric in   quasi-isotropic coordinates of the form given in (\ref{isotropic}) \cite{shibata,brandtseidel}. Let us define
\begin{eqnarray}
r_\mathrm{K} & = & r \left( 1 + \frac{m}{r} + \frac{m^2 - a^2}{4 r^2} \right), \\ 
\Delta_\mathrm{K} & = & r_\mathrm{K}^2 -2r_\mathrm{K}+a^2, \\
\Sigma_\mathrm{K} & = & r_\mathrm{K}^2 + a^2 \cos^2 \theta,
\end{eqnarray}
where $m$ and $a m$ denote the asymptotic mass and the angular momentum of the Kerr spacetime, respectively. The Kerr metric can be now written as
\begin{eqnarray}
g & = & - \alpha_\mathrm{K}^2 dt^2 + \psi_\mathrm{K}^4 e^{2q_\mathrm{K}} (dr^2 + r^2 d\theta^2) + \nonumber \\
&& \psi_\mathrm{K}^4 r^2 \sin^2 \theta (\beta_\mathrm{K} dt + d \varphi)^2,
\end{eqnarray}
where
\begin{eqnarray}
\psi_\mathrm{K} & = & \frac{1}{\sqrt{r}}\Bigl( r^2_\mathrm{K}  +a^2 +2ma^2\frac{r_\mathrm{K}\sin^2\theta  }{\Sigma_\mathrm{K}}\Bigr)^{1/4}, \\
\beta_\mathrm{K} & = & -\frac{2mar_\mathrm{K}}{(r^2_\mathrm{K}+a^2)\Sigma_\mathrm{K} +2ma^2r_\mathrm{K} \sin^2\theta}, \\
\alpha_\mathrm{K} & = & \left[ \frac{ \Sigma_\mathrm{K} \Delta_\mathrm{K}}{(r_\mathrm{K}^2+a^2)\Sigma_\mathrm{K}+2ma^2r_\mathrm{K} \sin^2\theta} \right]^{1/2}, \label{eq:alpha_K}\\
e^{q_\mathrm{K}} & = & \frac{\Sigma_\mathrm{K}}{\sqrt{(r^2_\mathrm{K}+a^2)\Sigma_\mathrm{K} +2ma^2r_\mathrm{K} \sin^2\theta}}.
\end{eqnarray}

We will apply the puncture formalism in the form presented in \cite{shibata}. Let $m$ and $a$ be parameters, corresponding to some Kerr spacetime. We define $r_\mathrm{s} = \frac{1}{2}\sqrt{m^2 - a^2}$; thus  for the Kerr metric with the asymptotic mass $m$ and the asymptotic angular momentum $am$, the event horizon coincides with a coordinate sphere $r = r_\mathrm{s}$.  In the general self-gravitating case, we replace the functions $\psi$ and $\alpha$ (the lapse) by $\phi$ and $B$ defined by the following relations
\begin{equation}
\label{puncture}
\psi = \left( 1 + \frac{r_\mathrm{s}}{r} \right) e^\phi, \quad \alpha \psi = \left( 1 - \frac{r_\mathrm{s}}{r} \right) e^{-\phi} B.
\end{equation}

The shift vector is split into two parts, $\beta = \beta_\mathrm{K} + \beta_\mathrm{T}$; their construction is described below. The   non-vanishing components of the extrinsic curvature of the slices of constant time $t$ can be expressed as
\begin{equation}
\label{Krf}
K_{r \varphi} = K_{\varphi r} = \frac{H_\mathrm{E} \sin^2 \theta}{\psi^2 r^2} +  \frac{1}{2 \alpha} \psi^4 r^2 \sin^2 \theta \partial_r \beta_\mathrm{T},
\end{equation}
\begin{equation}
\label{Ktf}
K_{\theta \varphi} = K_{\varphi \theta} = \frac{H_\mathrm{F} \sin \theta}{\psi^2 r} + \frac{1}{2 \alpha} \psi^4 r^2 \sin^2 \theta \partial_\theta \beta_\mathrm{T},
\end{equation}
where $H_\mathrm{E}$ and $H_\mathrm{F}$ are given by
\begin{eqnarray}
H_\mathrm{E} & = & \frac{ma \left[ (r_\mathrm{K}^2 - a^2) \Sigma_\mathrm{K} + 2 r_\mathrm{K}^2 (r_\mathrm{K}^2 + a^2) \right]}{\Sigma_\mathrm{K}^2}, \\
H_\mathrm{F} & = & - \frac{2 m a^3 r_\mathrm{K} \sqrt{\Delta_\mathrm{K}} \cos \theta \sin^2 \theta}{\Sigma_\mathrm{K}^2}.
\end{eqnarray}
Equations (\ref{Krf}) and (\ref{Ktf}) define $\beta_\mathrm{T}$. One can check that $\beta_\mathrm{T} = 0$ for the Kerr solution. In a sense, $\beta_\mathrm{K}$ is associated with the black hole, while $\beta_\mathrm{T}$ corresponds to the torus.

The Einstein equations can be written as the following system of equations for the functions $q$, $\phi$, $B$ and $\beta_\mathrm{T}$:
\begin{widetext}
\begin{subequations}
\label{main_sys}
\begin{eqnarray}
\left[ \partial_{rr} + \frac{1}{r } \partial_r  + \frac{1}{r^2} \partial_{\theta \theta}  \right] q & = & S_q, \label{47}\\
\left[ \partial_{rr} + \frac{2 r  }{r^2 - r_\mathrm{s}^2} \partial_r + \frac{1}{r^2} \partial_{\theta \theta} + \frac{  \cot{\theta}}{r^2}  \partial_\theta \right] \phi & = & S_\phi, \label{44} \\
\left[ \partial_{rr} + \frac{3 r^2 +  r_\mathrm{s}^2}{r(r^2 - r_\mathrm{s}^2)} \partial_r + \frac{1}{r^2} \partial_{\theta \theta} + \frac{2 \cot{\theta}}{r^2}  \partial_\theta \right] B & = & S_B, \label{45} \\
\left[ \partial_{rr} + \frac{4 r^2 - 8 r_\mathrm{s} r + 2 r_\mathrm{s}^2}{r(r^2 - r_\mathrm{s}^2)} \partial_r + \frac{1}{r^2} \partial_{\theta \theta} + \frac{3 \cot{\theta}}{r^2}  \partial_\theta \right]  \beta_\mathrm{T} & = & S_{\beta_\mathrm{T}}.  \label{46}
\end{eqnarray}
 \end{subequations}
Therein the source terms are given by
\begin{subequations}
\label{sources}
\begin{eqnarray}
S_q & = & -8 \pi e^{2q} \left( \psi^4 p - \frac{\rho h u_\varphi^2}{r^2 \sin^2 \theta} \right) + \frac{3 A^2}{\psi^8} + 2 \left[ \frac{r - r_\mathrm{s}}{r(r + r_\mathrm{s})} \partial_r + \frac{\cot \theta}{r^2} \partial_\theta \right] b + \left[ \frac{8 r_\mathrm{s}}{r^2 - r_\mathrm{s}^2} + 4 \partial_r (b - \phi) \right] \partial_r \phi \nonumber \\
& & + \frac{4}{r^2} \partial_\theta \phi \partial_\theta (b - \phi), \\
S_\phi & = & - 2 \pi e^{2q} \psi^4 \left( \rho_\mathrm{H} - p + \frac{\rho h u_\varphi^2}{\psi^4 r^2 \sin^2 \theta} \right) - \frac{A^2}{\psi^8} - \partial_r\phi \partial_r b - \frac{1}{r^2} \partial_\theta \phi \partial_\theta b - \frac{1}{2} \left[ \frac{r - r_\mathrm{s}}{r (r + r_\mathrm{s})} \partial_r b + \frac{\cot \theta}{r^2} \partial_\theta b \right], \\
S_B & = & 16 \pi B e^{2q} \psi^4 p, \\
S_{\beta_\mathrm{T}} & = & \frac{16 \pi \alpha e^{2q} j_\varphi}{r^2 \sin^2 \theta} - 8 \partial_r \phi \partial_r \beta_\mathrm{T} + \partial_r b \partial_r \beta_\mathrm{T} - 8 \frac{\partial_\theta \phi \partial_\theta \beta_\mathrm{T}}{r^2} + \frac{\partial_\theta b \partial_\theta \beta_\mathrm{T}}{r^2}.
\end{eqnarray}
\end{subequations}
\end{widetext}
The  function $\beta_\mathrm{K}$ satisfies the equation
\begin{equation}
\label{betak_eq}
\partial_r \beta_\mathrm{K} = 2 H_\mathrm{E} B e^{-8 \phi} \frac{(r - r_\mathrm{s})r^2}{(r + r_\mathrm{s})^7}.
\end{equation}
In the above formulas $B = e^b$ and
\begin{equation}
\label{a2formula}
A^2 = \frac{(\psi^2 K_{r \varphi})^2}{r^2 \sin^2 \theta} + \frac{(\psi^2 K_{\theta \varphi})^2}{r^4 \sin^2 \theta}.
\end{equation}
There appear also functions 
\begin{equation}
\rho_\mathrm{H} = \alpha^2 \rho h (u^t)^2 - p
\end{equation}
and
\begin{equation}
j_\varphi = \alpha \rho h u^t u_\varphi.
\end{equation}

There are imposed  boundary conditions at the   surface given by $r = r_\mathrm{s}$. They read
\begin{equation}
\partial_r q = \partial_r \phi = \partial_r B = \partial_r \beta_\mathrm{T} = 0.
\end{equation}
Equation (\ref{46}) requires a more stringent boundary condition. Following \cite{shibata} we set $\beta_\mathrm{T} = O[(r - r_\mathrm{s})^4]$, which is equivalent to $\beta_\mathrm{T} = \partial_r \beta_\mathrm{T} = \partial_{rr} \beta_\mathrm{T} = \partial_{rrr} \beta_\mathrm{T} = 0$ at $r = r_\mathrm{s}$.

One can  show, with the preceding conditions, that the two-surface $r = r_\mathrm{s}$ embedded in a hypersurface of constant time $\Sigma_t$ is a minimal surface. 

We will refer to the system of equations (\ref{rot_law_eq}), (\ref{bernoulli2}), (\ref{main_sys}), and (\ref{betak_eq}) as the Einstein-Euler equations.

\subsection{Mass and angular momentum}

Black hole-torus systems are characterized   by   masses and angular momenta of their constituents. The Arnowitt-Deser-Misner (ADM) asymptotic mass of the whole system is an obvious choice. It is defined as an asymptotic surface integral, but we choose  to compute the ADM mass using an equivalent formula \cite{shibata}:
\begin{eqnarray} 
M_\mathrm{ADM} = \sqrt{m^2 - a^2} + M_1, 
\end{eqnarray}
where
\begin{eqnarray} 
M_1 = - 2 \int_{r_\mathrm{s}}^\infty dr \int_0^{\pi/2} d \theta (r^2 - r_\mathrm{s}^2) \sin \theta S_\phi, 
\end{eqnarray}
and $m$ is the black-hole mass parameter introduced in Sec.\ \ref{einstein_equations}.

The quasilocal mass of the black hole is given by the   commonly used formula  
\begin{eqnarray}
 M_\mathrm{BH} = M_\mathrm{irr} \sqrt{1 + \frac{J_\mathrm{H}^2}{4 M_\mathrm{irr}^4}}.
\label{Christodoulou}
\end{eqnarray}

The mass of the disk is defined as \cite{shibata}
 \begin{eqnarray}
 M_\mathrm{disk} = 4\pi \int_{r_\mathrm{s}}^{\infty} r^2 dr \int_{0}^{\pi/2} \rho \alpha u^t \psi^6 e^{2q} \sin\theta d\theta.
\label{disk_mass}
\end{eqnarray}

$J_\mathrm{H}$ is the angular momentum of the black hole,
\begin{eqnarray}
J_\mathrm{H} = \frac{1}{4} \int_0^{\pi/2} d \theta \left( \frac{r^4 \sin^3 \theta \psi^6 \partial_r \beta}{\alpha} \right)_{r=r_\mathrm{s}}, 
\end{eqnarray}
and $M_\mathrm{irr}$ denotes the so-called irreducible mass, defined as
\begin{eqnarray}
M_\mathrm{irr} = \sqrt{\frac{A_\mathrm{H}}{16 \pi}},
\end{eqnarray}
where $A_\mathrm{H}$ is the area of the horizon,
\begin{eqnarray}
A_\mathrm{H} = 4 \pi \int_0^{\pi/2} d \theta \left( \psi^4 e^q r^2 \sin \theta \right)_{r=r_\mathrm{s}}.
\end{eqnarray}
This definition is inspired by  the observation of Christodoulou \cite{christodoulou} that for the Kerr solution the asymptotic mass is equal to the right-hand side of (\ref{Christodoulou}). In the presence of matter that satisfies standard energy conditions, we have $M_\mathrm{ADM}\ne M_\mathrm{BH}$.

The angular momentum of the torus is defined as
\begin{eqnarray}
J_1 & = & \int \sqrt{- g} T\indices{^t_\varphi} d^3 x \nonumber \\
& = & 4 \pi \int_{r_\mathrm{s}}^\infty dr \int_0^{\pi/2} d\theta r^2 \sin \theta \alpha \psi^6 e^{2q} \rho h u^t u_\varphi.
\end{eqnarray}
The above definition follows from the conservation law $\eta^\nu \nabla_\mu T\indices{^\mu_\nu} = \nabla_\mu (T\indices{^\mu_\nu} \eta^\nu) = 0$ \cite{leshouches} for the Killing vector $\eta^\mu = (0,0,0,1)$. The total angular momentum can be expressed as
\begin{equation}
J = J_\mathrm{H} + J_1.
\end{equation}

We point out that the value assigned to $J_\mathrm{H}$ depends on the assumed boundary conditions at $r = r_\mathrm{s}$. In our case (and in \cite{shibata}) the condition $\partial_r \beta_\mathrm{T} = 0$ at $r = r_\mathrm{s}$ yields $J_\mathrm{H} = a m$. Obvioulsy, the  natural definition of the black-hole spin would be
\begin{equation}
\hat a = \frac{J_\mathrm{H}}{M_\mathrm{BH}} = \frac{a m}{M_\mathrm{BH}}.
\end{equation}
In general $\hat a \neq a$, but we have equality for a massless disk.

The discussion of other mass measures (including a quasilocal toroidal mass)  and  relations between them can be found in \cite{shibata,mgfop}.

We should  stress out that in the presence of self-gravitating disks the    mass of the black hole $M_\mathrm{BH}$ is larger than  the black hole mass parameter   $m$. The effect is negligible for light tori (in vacuum obviously   $M_\mathrm{BH}=m$) but it becomes noticeable for heavy disks. This is true assuming  a ``reasonable'' matter; the perfect fluid considered in this paper, with the DD2-Timmes-Swesty equation of state, is ``reasonable''.

\section{Tabulated equations of state  }
\label{numerics}

In this paper we use a tabulated equation of state after \cite{sho_winter,sho_autumn}. This equation of state is based on two prescriptions: the Density-Dependent DD2 model in the high-density sector and Timmes and Swesty equation of state for low values of density \cite{equation_of_state}. For this equation of state the specific enthalpy $h$, the energy density $\epsilon$ and the pressure $P$ are the functions of the density $\rho$, the electron fraction $Y_\mathrm{e}$, and the temperature $T$. In order to compute values of $\rho$, $Y_\mathrm{e}$, and $T$ from $h$, one has to assume two relations between these variables. The first was already made during derivation of the Bernoulli equation (\ref{bernoulli3})---that the specific entropy (entropy per baryon) $s$ is constant. The second assumption is that $Y_\mathrm{e}$ is a function of $\rho$ only.  Here we define $Y_\mathrm{e}$ after \cite{sho_winter,sho_autumn}. It ranges from $0.5$ to $0.07$ and depends on $\rho$ as follows: for $\rho \lessapprox 10^7~\mathrm{g}/\mathrm{cm}^3$, $Y_\mathrm{e}=0.5$; for $\rho \gtrapprox 10^{11}~\mathrm{g}/\mathrm{cm}^3$, $Y_\mathrm{e}=0.07$; for $\rho\in(10^7,10^{11})~\mathrm{g}/\mathrm{cm}^3$, $Y_\mathrm{e}$ decreases linearly from $0.5$ to $0.07$.

The minimal value of the specific enthalpy in the disk $h_\mathrm{min}$ is lower than $h=c^2$ due to the presence of the nuclear binding energy. One has to choose $h_\mathrm{min}$ in order to solve Eq.\ (\ref{bernoulli2}) at the disk edges in the equatorial plane, which is necessary for the computation of the angular velocity $\Omega$. Here we took   $h_\mathrm{min}\approx 0.9987$, the lowest value in the table, which corresponds to the rest-mass density $\rho_\mathrm{min}\in(0.1,0.6)~\mathrm{g}/\mathrm{cm}^3$. The value of $h_\mathrm{min}$ varies slightly for different equations of state (with different values of entropy per baryon $s$ and/or the electron fraction $Y_\mathrm{e}$). In order to calculate hydrodynamic quantities from the tabulated equation of state, we used for each quantity an interpolation linear in logarithms of $(h-h_\mathrm{min})/c^2$, $\rho$, $P$, $kT$ (where $k$ is Boltzmann constant). Because of  numerical problems with the interpolation we subtracted from each appropriate tabular value a small number  $10^{-16}$ while setting values of $h_\mathrm{min} $. This allows us to interpolate all points of the table.

\section{  Rotation laws of \cite{sho_winter} versus \cite{KM2020}} 
\label{rotation}

Fujibayashi et al.\ \cite{sho_winter} have assumed the rotation law 
\begin{equation}
\tilde j= A_j\Omega^{\delta }
\label{sho}
\end{equation}
 that is similar to the rotation law (\ref{momentum}) in some aspects. We shall investigate in this Section  these similarities.
 
The first observation is that both rotation laws have the same Newtonian limit. Indeed, in the Newtonian limit $c\rightarrow \infty $ one gets from (\ref{sho})  $\tilde j= \hat A_j\Omega^{\delta }$, where $\hat A_j=\lim_{c\rightarrow \infty}A_j$; this  implies, using  (\ref{rot_law_eq}),  the   angular velocity $\Omega = \frac{\left( \hat A_j\right)^{1/\left( 1-\delta \right)}}{r^{2/(1-\delta )}}$. On the other hand, formulae  (\ref{momentum}) and (\ref{rot_law_eq}) give in the Newtonian limit  $\Omega = \frac{\hat w}{r^{2/(1-\delta )}}$ \cite{jmmp}. Here $\hat w=\lim_{c\rightarrow \infty}w$. Thus these two Newtonian limits coincide if $\hat w=\hat A_j^{1/(1-\delta )}$. 

In the second step, assume the rotation law given by Eq.\ (\ref{sho}); one gets the Euler-Bernoulli equation in the form
\begin{equation}
\label{aaaa}
h \sqrt{\alpha^2 - \psi^4 r^2 \sin^2 (\Omega + \beta)^2} + \frac{A_j}{1 + \delta} \Omega^{1 + \delta} = C_1.
\end{equation}
It is easy to see that the above relation is in fact a special case of Eq.\ (\ref{bernoulli3}). Indeed, let us take $a_\mathrm{rot} = 0$ and $\kappa = 1/(1 + \delta)$ in Eq.\ (\ref{bernoulli3}). This yields
\begin{equation}
\label{aaab}
h \sqrt{\alpha^2 - \psi^4 r^2 \sin^2 (\Omega + \beta)^2} = C^\prime - \frac{C^\prime w^{1 - \delta}}{1 + \delta} \Omega^{1+\delta}.
\end{equation}
Clearly, Eqs.\ (\ref{aaaa}) and (\ref{aaab}) coincide, provided that $C_1 = C^\prime$ and $A_j = C^\prime w^{1 - \delta}$.

 There are two useful relations between $j$ and $\tilde j$. 
One gets
\begin{equation}
\label{jj}
    \frac{\tilde j}{j} = \frac{h}{u^t}
\end{equation}
directly from definitions  $j=u^tu_\varphi$ and $\tilde j=hu_\varphi$.
Employing now   (\ref{bernoulli4})   
\begin{equation}
\frac{h}{u^t} + \int \tilde j(\Omega) d\Omega = C_1,
\end{equation}
we arrive at 
\begin{equation}
    \frac{\tilde j}{j} + \int \tilde j(\Omega) d\Omega = C_1.
\end{equation} 
Thus
\begin{equation}
   \frac{ \tilde j}{ j} =  C_1 - \int \tilde j(\Omega) d\Omega .
\end{equation}
In a similar vein Eqs.\ (\ref{bernoulli3}) and (\ref{jj}) yield
\begin{equation}
    \ln\left(\frac{\tilde j}{j}\right) + \int j(\Omega) d \Omega = C.
\end{equation}

\section{On parametrization of solutions and numerical procedure }

\subsection{Parametrization of solutions}

The black hole-torus system is described by Eqs.\ (\ref{momentum}), (\ref{rot_law_eq}), (\ref{bernoulli2}), (\ref{main_sys}), (\ref{betak_eq}) and the tabulated equation of state (cf.\ Sec.\ \ref{numerics}).

In our calculations we choose the mass parameter $m$ to be a unit of mass and length, so $m=1$. We assume further that it corresponds to three solar masses, $3M_\odot$. We remarked earlier that the quasilocal mass of the black-hole $M_\mathrm{BH}$ in the presence of the disk can be larger  than $m$. In most cases reported below  the difference doesn't exceed $1\%$ and it   approaches $10\%$ only in solutions with disks having masses of the order of $m$. The inner and outer coordinate radii of disks at the symmetry plane $\theta =\pi /2$ are  $r_1=2$ and $r_2=40$ (with one exception) respectively; in SI units they  are 9 km or 180 km.

The numerical method used in this paper is a modification of a scheme described and tested in \cite{kkmmop2}. One of the changes with respect to the version described in \cite{kkmmop2} is the implementation of a very efficient PARDISO linear algebra library \cite{pardiso}, which is now used instead of LAPACK \cite{lapack}. Major changes were enforced by the implementation of a new rotation law and, more importantly, the tabulated equation of state.

A version described in \cite{kkmmop2} used polytropic equations of state of the form
\[ p = K \rho^\Gamma, \]
where $K$ and $\Gamma$ are constant. Solutions were specified by setting $m$, $a$, $r_1$, $r_2$, $\Gamma$, and the maximum value of the rest-mass density within the disk $\rho_\mathrm{max}$. The rotation law was prescribed up to a constant, corresponding roughly to $w$ in Eqs.\ (\ref{momentum}), (\ref{keplerian_rl}), or (\ref{kmm}). This meant, in particular, that the value of the polytropic constant $K$ was not specified, but computed from the requirement concerning $\rho_\mathrm{max}$. We found this scheme to be much more effective than simply fixing $K$ a priori and computing $\rho_\mathrm{max}$ as a part of the solution. On the other hand, with the tabulated equation of state, this point requires a change. We now specify a constant value of the specific entropy $s$ (in the units of the Boltzmann constant $k$) together with a relation between the electron fraction $Y_\mathrm{e}$ and the rest-mass density, as described in Sec.\ \ref{numerics}. With these settings, the equation of state becomes essentially barotropic---the relation between the rest-mass density and the pressure (or the specific enthalpy) becomes fixed. As a consequence, there is no freedom in specifying $\rho_\mathrm{max}$ as a parameter, which, as we shall see, is an obstacle in obtaining solutions corresponding to massive disks. In summary, solutions with light disks are specified by choosing the equation of state (fixing the value of $s$) and the parameters $m$, $a$, $r_1$, $r_2$, $a_\mathrm{rot}$, $\delta$, and $\kappa$. That means, as before, that the rotation law is prescribed up to the parameter $w$, which has to be computed as a part of the solution, together with the constant $C^\prime$ in Eq.\ (\ref{bernoulli2}).

The key observation is that, similarly to the situation described in \cite{dyba}, the solutions are not unique with respect to the above parametrization. Given fixed parameters $m$, $a$, $r_1$, $r_2$, $a_\mathrm{rot}$, $\delta$, $\kappa$, and the equation of state, one can still obtain two distinct solutions, differing in the total asymptotic mass (or the mass of the disk). There is a branch of solutions corresponding to relatively light disks, which we denote as branch I, and a branch of solutions corresponding to more massive disks, referred to as branch II. Numerical procedures used to obtain solutions belonging to these two branches are slightly different; they are described in the next subsection.

\subsection{Finding solutions corresponding to light and massive disks}

Solutions corresponding to light disks (branch I) are obtained by an iterative procedure, in which each iteration starts with a computation of the angular velocities $\Omega_1$ and $\Omega_2$ at the inner, $(r,\theta) = (r_1,\pi/2)$, and outer, $(r,\theta) = (r_2,\pi/2)$, edges of the disk. This is done by a Newton-Raphson method, assuming Eqs.\ (\ref{rot_law_eq}) and (\ref{bernoulli2}), and the rotation law (\ref{momentum}). We assume that the edges of the torus correspond to $h = h_\mathrm{min}$, as described in Sec.\ \ref{numerics}. This procedure also yields the values of constants $w$ and $C^\prime$. In the subsequent step we compute, from Eq.\ (\ref{rot_law_eq}), the values of $\Omega$ in all grid points within the disk. Next, Eq.\ (\ref{bernoulli2}) is used to determine the specific enthalpy $h$ in the torus. Given $h$, we compute the rest-mass density $\rho$ and the pressure $p$ from the tabulated equation of state (for a fixed value of the specific entropy $s$), interpolating linearly in logarithms of $h$, $\rho$ and  $p$ as described in Sec.\ \ref{numerics}. The iteration is concluded with solving the Einstein equations (\ref{main_sys}) and (\ref{betak_eq}) for functions $q$, $\phi$, $B$, $\beta_T$, and $\beta_K$, which amounts to the main computational cost of the entire method.

The procedure used for finding solutions corresponding to massive disks is different. The trick is to restore (temporarily and only in a very specific sense) a control of $\rho_\mathrm{max}$. Each iteration proceeds as before, until a point at which a new distribution of the specific enthalpy is found, as follows from Eq.\ (\ref{bernoulli2}). We then search for the maximum of $h$ within the disk and, using the tabulated equation of state, find a corresponding value of the rest-mass density $\rho_0$. This allows us to introduce an auxiliary parameter $\lambda = \rho_\mathrm{max}/\rho_0$, where $\rho_\mathrm{max}$ is a desired maximal value of $\rho$. In the next step we construct a rescaled tabulated equation of state by setting $\tilde \rho = \lambda \rho$, $\tilde p = \lambda p$. The idea behind this choice is that for barotropic equations of state the specific enthalpy is given by $h = \int dp/\rho$, and thus the above scaling leaves $h$ unchanged. This rescaled equation of state is used to compute the rest-mass density and the pressure within the disk. The remainder of each iteration proceeds as before with a solution of Einstein equations (\ref{main_sys}) and (\ref{betak_eq}). These iterations are repeated until a certain level of convergence is reached, which also means that the parameter $\lambda$ converges to a fixed value. Of course, the solution obtained in this way usually corresponds to an unphysical, rescaled equation of state. Solutions corresponding to massive disks are obtained by changing $\rho_\mathrm{max}$ every few thousands of iterations, until finally they converge to a solution with $\lambda = 1$ (with the accuracy $\lambda - 1 \le 10^{-8}$), obeying the original, physical equation of state. More precisely, every 1000 iterations $\rho_\mathrm{max}$ is increased, if $\lambda > 1$ and decreased, if $\lambda < 1$. The opposite choice, i.e., decreasing $\rho_\mathrm{max}$ for $\lambda>1$ and increasing for $\lambda<1$, can be used to retrieve solutions corresponding to light tori. Unfortunately, the whole procedure is quite sensitive to the initial choice of $\rho_\mathrm{max}$---the solution may diverge, if $\rho_\mathrm{max}$ is significantly different from the target value. Therefore, we always chose initial $\rho_\mathrm{max}$ to be not smaller than the value of $\rho_\mathrm{max}$ on the corresponding light branch and of the order of magnitude not larger than that of the targetted value of $\rho_\mathrm{max}$.

\subsection{Numerical tests}
\label{sec:Vtest}

\begin{figure}[ht]
\includegraphics[width=1\columnwidth]{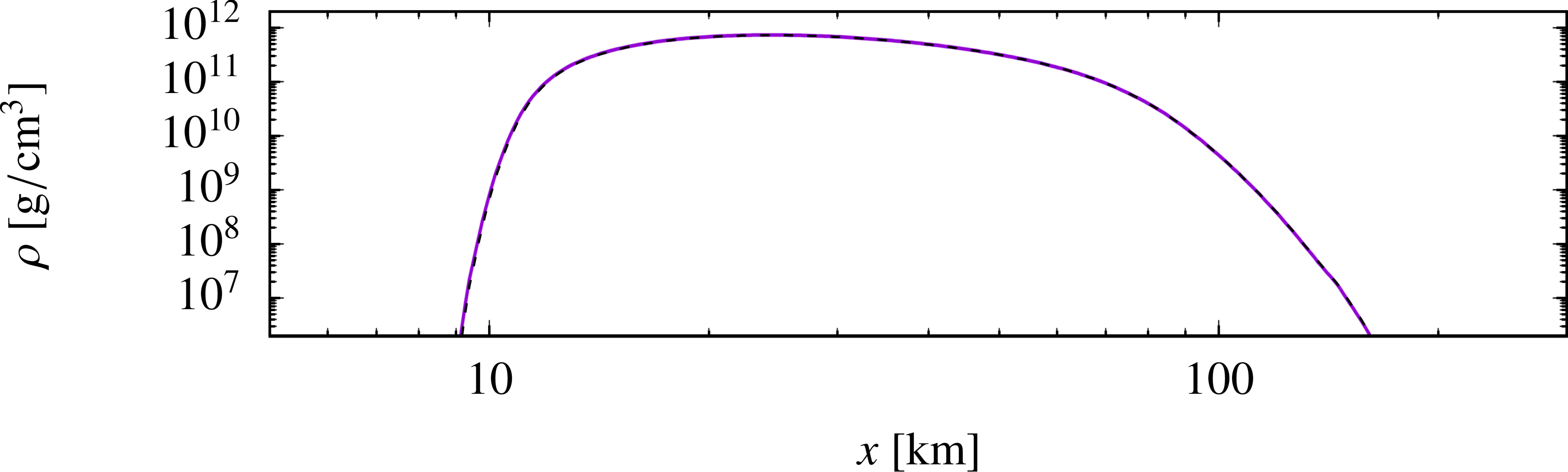}
\caption{A comparison between the equatorial ($x = r \sin \theta$) rest-mass density profile shown in Fig.\ 1 of \cite{sho_winter} (violet line) and the corresponding profile computed with the present code, using the rotation law $j_\mathrm{Sh}$ (dashed line). The inner and outer coordinate radii of the disk are $r_1=2$, $ r_2=41$, respectively. We assume $m = 1$, $a = 0.8$, and $\delta=-1/7$. The specific entropy in the disk $s = 6k$. The mass of the disk is $M_\mathrm{disk}=0.1M_\odot$.}
\label{fig:rho_numerics}
\end{figure}

Fujibayashi  et al.\ constructed  in \cite{sho_winter,sho_autumn} stationary solutions assuming the DD2-Timmes-Swesty equation of state and the rotation law  $\tilde j(\Omega) =hu_\varphi =A_j\Omega^{\delta}$.
We reproduce  their solutions, using the relation described in Sec.\ \ref{rotation} between their and our approach.   
 We shall use  $j=u^tu_\varphi $. Choosing  in Eq.\ (\ref{momentum}) constants $a_\mathrm{rot}=0, \kappa = 1/(1+\delta )$, we get the rotation law $j_\mathrm{Sh}$.  

\begin{figure}[ht]
\includegraphics[width=1\columnwidth]{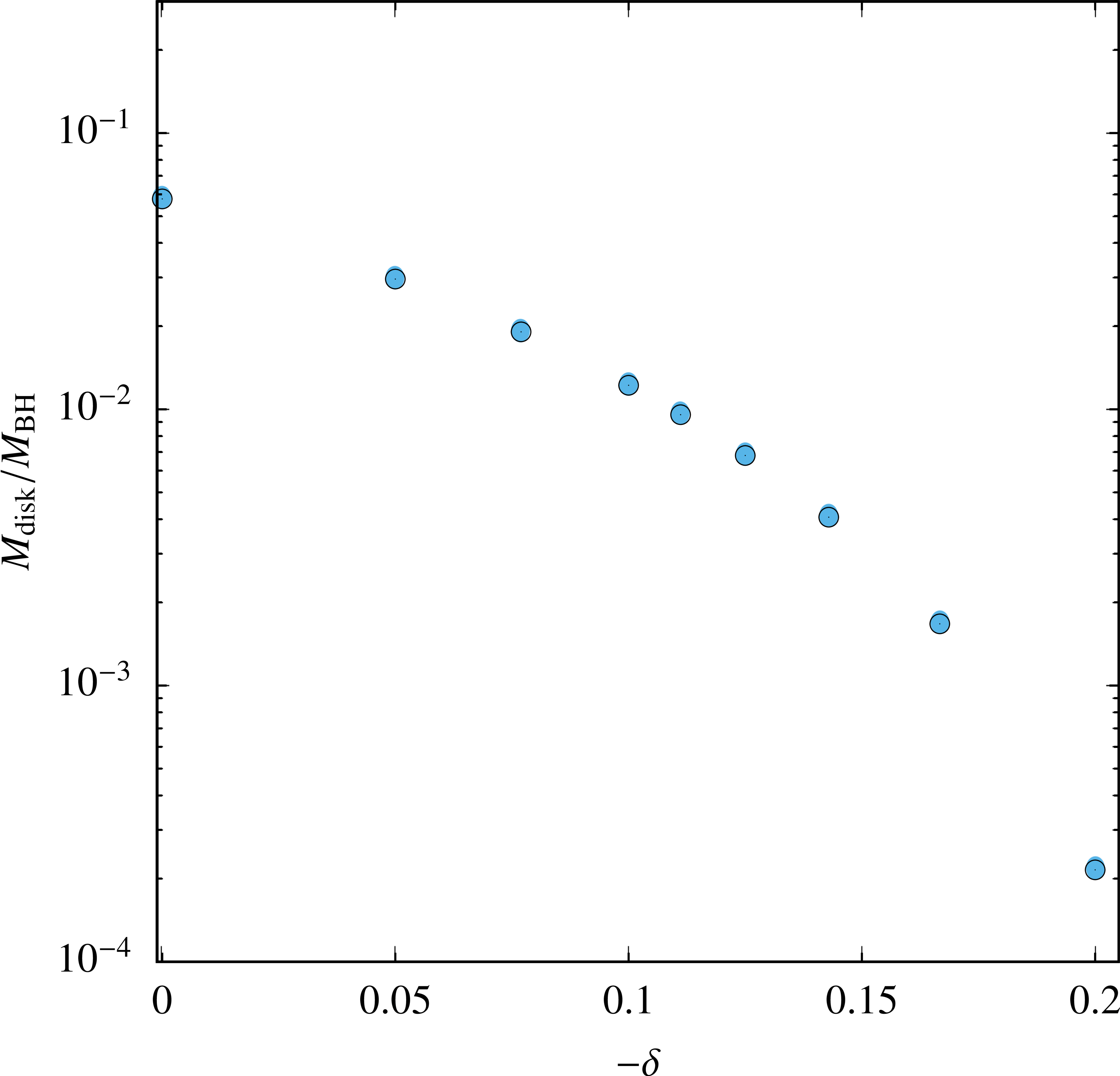}
\caption{A comparison between the ratios $M_\mathrm{disk}/M_\mathrm{BH}$ (relative disk masses) depicted in Fig.\ 2 of \cite{sho_winter} (blue dots) and those computed with the present code, assuming the rotation law $j_\mathrm{Sh}$ (empty circles). The inner and outer coordinate radii are $r_1 = 2$ and $ r_2 = 40$, respectively. The spin parameter $a=0.8$, and the specific entropy $s=8k$. The exponent $\delta$ in the rotation law changes from $-0.2$ to $0$.}
\label{fig:mdisk_numerics}
\end{figure}

Results taken from \cite{sho_winter} and those reproduced by the use of our procedure and $j_\mathrm{Sh}$, shall be compared in Figs.\ \ref{fig:rho_numerics} and \ref{fig:mdisk_numerics}. Strictly speaking, we shall deal with branch I of configurations with light disks.  

In Figure \ref{fig:rho_numerics} we compare the disk density profiles on the plane $\theta =\pi /2$ obtained in \cite{sho_winter} with the result of our calculation. Here $\tilde j(\Omega) = A_j\Omega^{-1/7}$. It is clear that the density profiles of \cite{sho_winter} and that of the present work, do agree. 

\begin{figure}[ht]
\includegraphics[width=1\columnwidth]{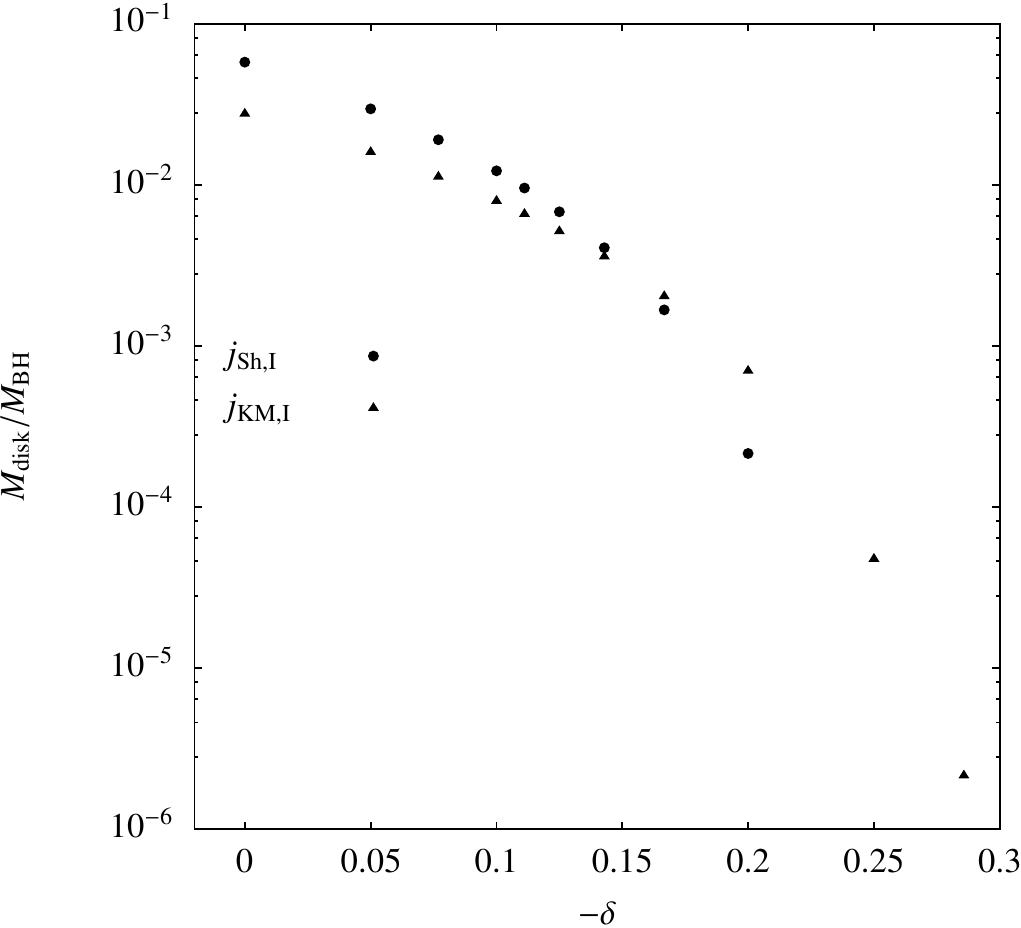}
\caption{The ratio $M_\mathrm{disk}/M_\mathrm{BH}$ vs.\ the parameter $\delta $ for two rotation laws---$j_\mathrm{Sh}$ and $j_\mathrm{KM}$. The graph shows data corresponding to the light disks (branch I). The parameters $r_1$, $ r_2$, $m$, $a$, and $s$ are the same as in Fig.\ \ref{fig:mdisk_numerics}. The exponent $\delta$ in the rotation law (\ref{kmm}) changes from $0$ to $-2/7$; there are no solutions in the case of $j_\mathrm{Sh}$ for $\delta =-1/4, -2/7$. For the rotation law $j_\mathrm{KM}$ the parameter $a_\mathrm{rot}=0.8$.}
\label{fig:mdisk_compare_I}
\end{figure}

 \begin{figure}[ht]
\includegraphics[width=1\columnwidth]{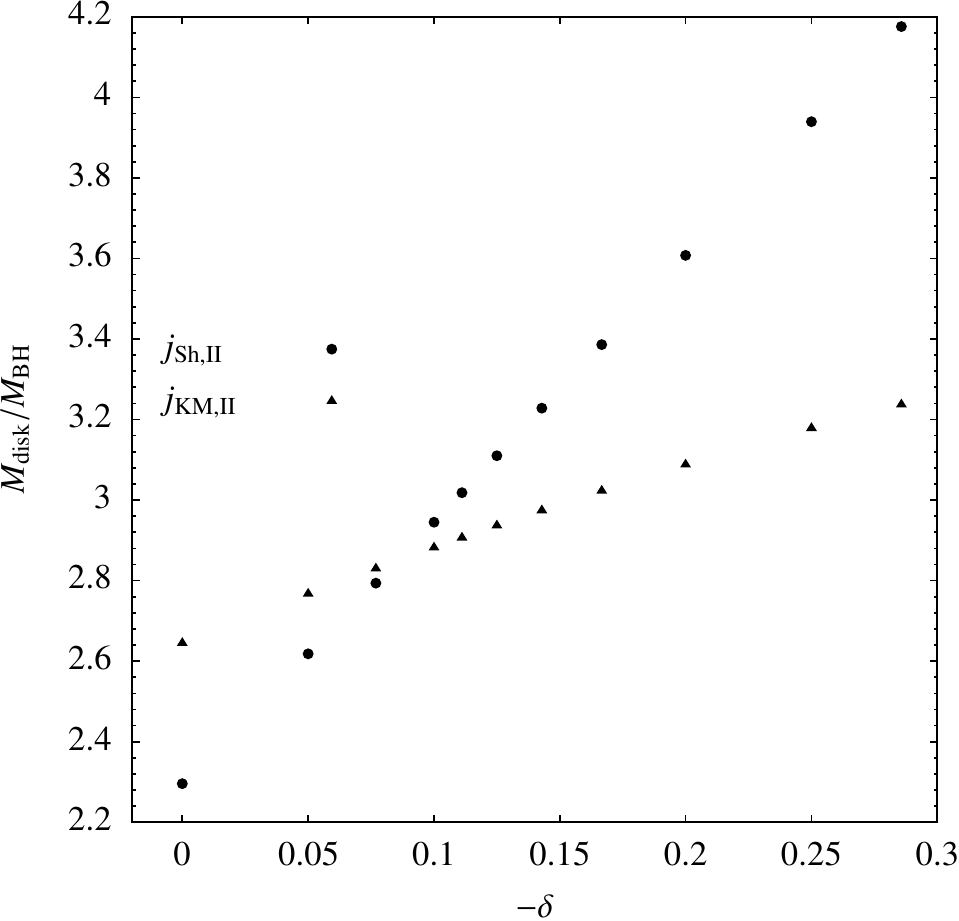}
\caption{The ratio $M_\mathrm{disk}/M_\mathrm{BH}$ vs.\ the parameter $\delta $ for two rotation laws---$j_\mathrm{Sh}$ and $j_\mathrm{KM}$. The graph shows data corresponding to heavy disks (branch II). The remaining parameters are as in  Fig.\ \ref{fig:mdisk_compare_I}.}
\label{fig:mdisk_compare_II}
\end{figure}

Figure \ref{fig:mdisk_numerics} shows masses of disks obtained for different values of the parameter $\delta $, $-0.2\le \delta \le 0$, assuming the rotation law of \cite{sho_winter}: $ j_\mathrm{Sh}  $.   Again we can conclude   that results of \cite{sho_winter} and those obtained by us  are essentially the same.  
 
\subsection{Solutions: $j_\mathrm{KM}$ versus $j_\mathrm{Sh}$}

In the next  two figures \ref{fig:mdisk_compare_I} and \ref{fig:mdisk_compare_II} we demonstrate how the relative masses (of disks versus the black holes) depend on the   parameter $\delta  $, within a shown interval. It is notable, that the aforementioned two branches of solutions, I and II, exist for both rotation laws $j_\mathrm{Sh}$ and $j_\mathrm{KM}$. Herein we put the specific entropy $s = 8 k$ and for $j_\mathrm{KM}$ we set $a_\mathrm{rot}=a$. The masses of disks corresponding to the two rotation laws  are roughly equal at $\delta \approx -0.15$ at the light branch and at  $\delta \approx -0.08$ at the heavy branch. They behave differently with the change of $\delta $, depending on the branch. In the case of light disks, shown in Fig.\ \ref{fig:mdisk_compare_I}, the mass decreases with $|\delta |$ and the falloff is faster for the rotation law $ j_\mathrm{Sh} $.  In the case of heavy disks, depicted in Fig.\ \ref{fig:mdisk_compare_II}, the mass increases with $|\delta |$ and the growth is faster for the rotation law $j_\mathrm{Sh}$.

\section{Bifurcation of solutions  } 
\label{bifurcation}

\begin{figure}[ht]
\includegraphics[width=1\columnwidth]{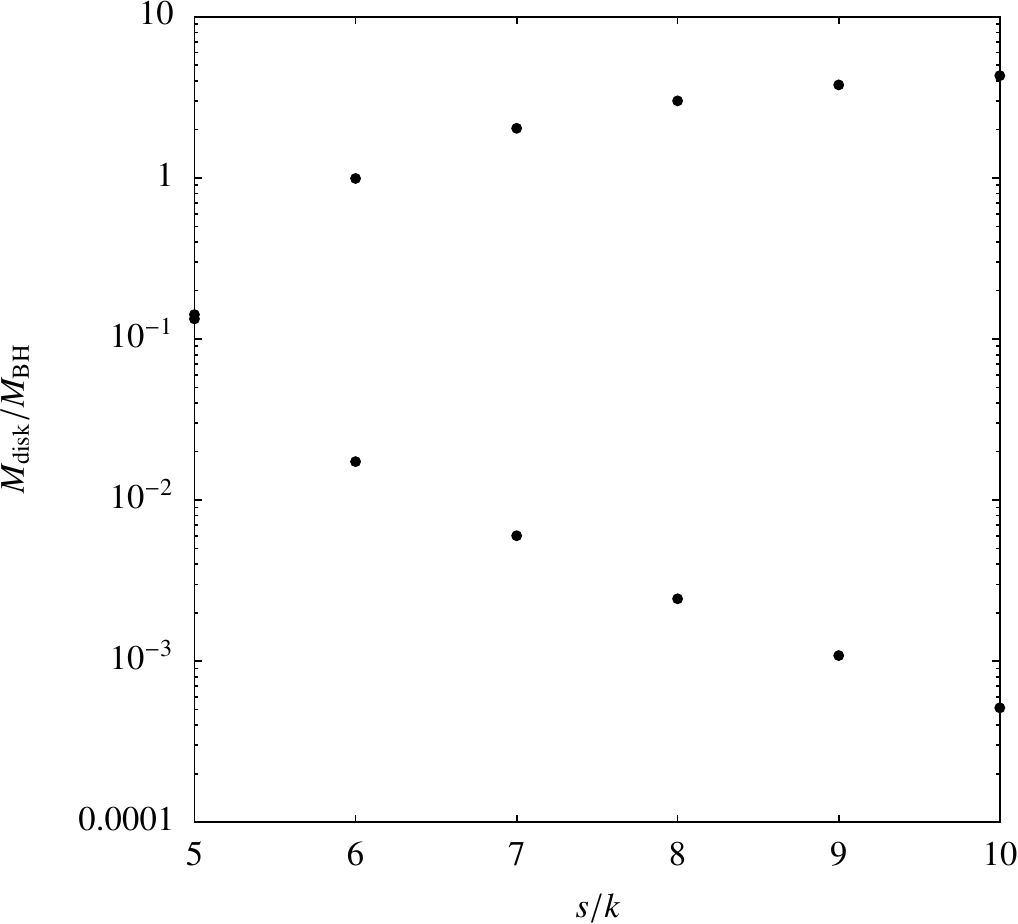}
\caption{The ratio $M_\mathrm{disk}/M_\mathrm{BH}$ vs.\ the specific entropy $s$ for the rotation law $j_\mathrm{KM}$. The inner and outer coordinate radii of disks are $r_1 = 2$, $r_2 = 40$. The mass parameter $m=1$, $a_\mathrm{rot}=0.8$, and the spin parameter $a=0.8$. Here $\delta^* = -0.1595$, and $M_\mathrm{disk}^* \in \left(0.40, 0.43\right) M_\odot$.}
\label{fig:Bifurcation_s}
\end{figure}    

In this section we shall work exclusively with the rotation law
$j_\mathrm{KM} $ of Eq.\ (\ref{kmm}), assuming $a_\mathrm{rot}=a$. We have studied a few dozens of solutions corresponding to different pairs $(\delta, s)$ consisting of the exponent in the rotation law and the specific entropy, respectively. Typically, for a chosen pair $(\delta ,s )$  there exist two  solutions---two configurations that differ in the mass ratio $M_\mathrm{disk}/M_\mathrm{BH}$. We have found that, if one of the parameters, $\delta$ or $s$, is fixed then the set of solutions consists of two branches, in which solutions are labeled  by the other parameter. We shall analyse the structure of this set in what follows. 
 
 \begin{figure}[ht]
\includegraphics[width=1\columnwidth]{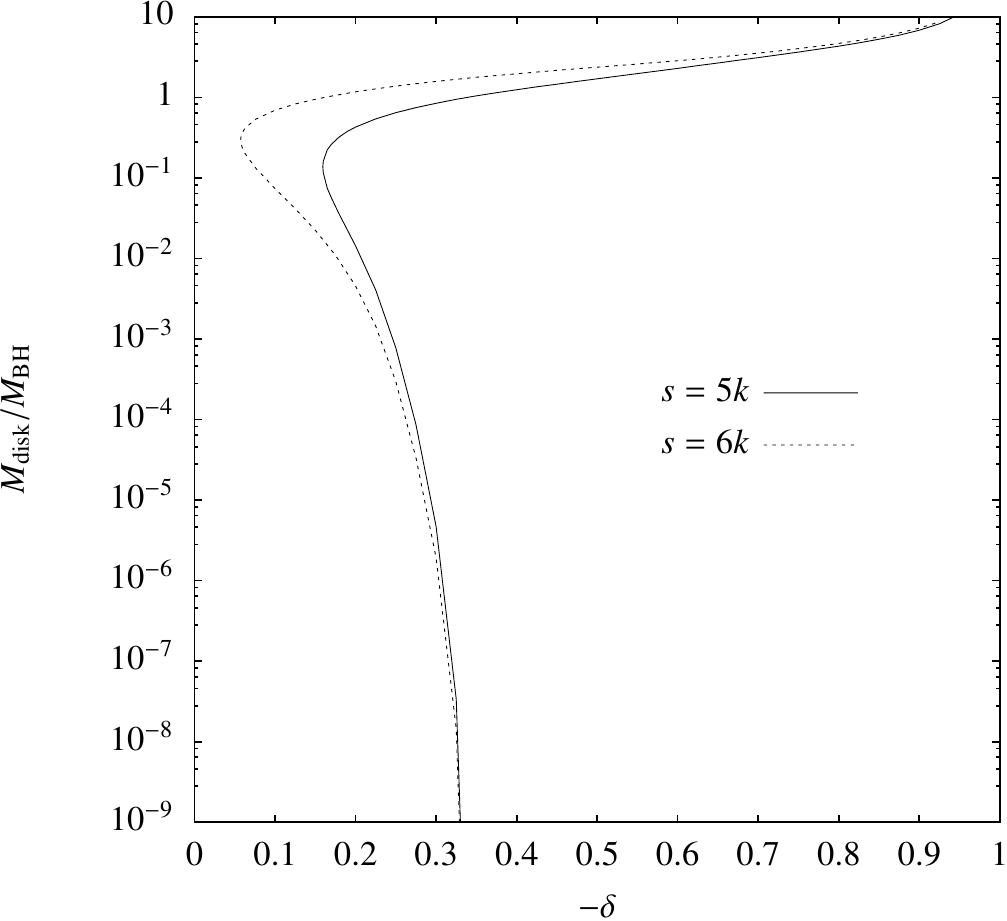}
\caption{The ratio $M_\mathrm{disk}/M_\mathrm{BH}$ vs.\ the parameter $\delta $ for the rotation law $j_\mathrm{KM}$. The parameters $m, r_1, r_2$,  $a_\mathrm{rot}$, and $a$ are the same as in Fig.\ \ref{fig:Bifurcation_s}. The specific entropy $s = 5 k$ (solid line) or $s = 6 k$ (broken line). The values of $\delta$ and disk masses corresponding to critical solutions are: $ \delta^*_1 \approx -0.1595$ and $M_\mathrm{disk}^*\in \left(0.40, 0.43\right)M_\odot$ (solid line); $ \delta^*  \approx -0.0576$ and $M_\mathrm{disk}^*\in \left(0.86,0.93\right)M_\odot$ (broken line).}
\label{fig:Bifurcation_lin}
\end{figure}

In our first study we fix $\delta\equiv \delta^*  =-0.1595$ and change the specific entropy parameter $s$. We have  found a sequence of solutions, shown in Fig.\ \ref{fig:Bifurcation_s}, that suggests the existence of a  bifurcation point. The specific entropy $s$ plays a role of the bifurcation parameter---two branches of solutions seem to originate  from the critical point $s=s^*= 5 k$. There exists a (critical) solution at the conjectured bifurcation point $s^*$ with a disk mass $M_\mathrm{disk}^*\in (0.40, 0.43 M_\odot )$. 
 
There is a trivial but serious restriction preventing further investigation that would require going into smaller values of $s$. We are  limited by the known to us tabulated equation of state, in which the smallest value of the  specific entropy is $s = 5 k$ and the set of values of $s/k$ is discrete: $(s/k)=(5,6,7,8,9,10)$. 
For that reason we are not able to investigate in more detail the neighbourhood of the above bifurcation point. For the same reason we could not resolve   an interesting issue, whether there exists a  subcritical solution, i.e., with subcritical values of the specific entropy parameter $s < 5 k$. Therefore we decided to study a different situation, with the specific entropy parameter $s$ being fixed and the exponent $\delta$ being the bifurcation parameter. We have  found two examples of bifurcation. In one case we have $s = 5 k$ and  the critical value of the bifurcation parameter $\delta^*_1 \approx-0.1595$. In the other case the specific entropy is $s=6 k$ and the critical exponents is $\delta^*_2\approx-0.0576$.
The relevant graphs are shown in Fig.\  \ref{fig:Bifurcation_lin}. 
  
In each case  the two branches of solutions, I and II  (with light and heavy disks, respectively), converge to a common vertex (the bifurcation point) when $\delta\rightarrow \delta^*_1$ or  $\delta\rightarrow \delta^*_2$. We have at the bifurcation points solutions corresponding to disks with masses $M_{\mathrm{disk},1}^*\in \left(0.40, 0.43\right)M_\odot$ (solid line) and  $M_{\mathrm{disk},2}^*\in \left(0.86, 0.93\right)M_\odot$ (broken line), respectively. The masses $M^*_{\mathrm{disk},1}$ and  $M^*_{\mathrm{disk},2}$ for critical solutions are not determined exactly due to the numerical difficulties that are typical for bifurcation; the time needed to generate a solution grows quickly, from one hour to a couple of days, when $\delta $ approaches a critical value. A more precise calculation would require a numerical run extending for a couple of weeks on a personal computer.
 
\begin{figure}[ht]
\includegraphics[width=1\columnwidth]{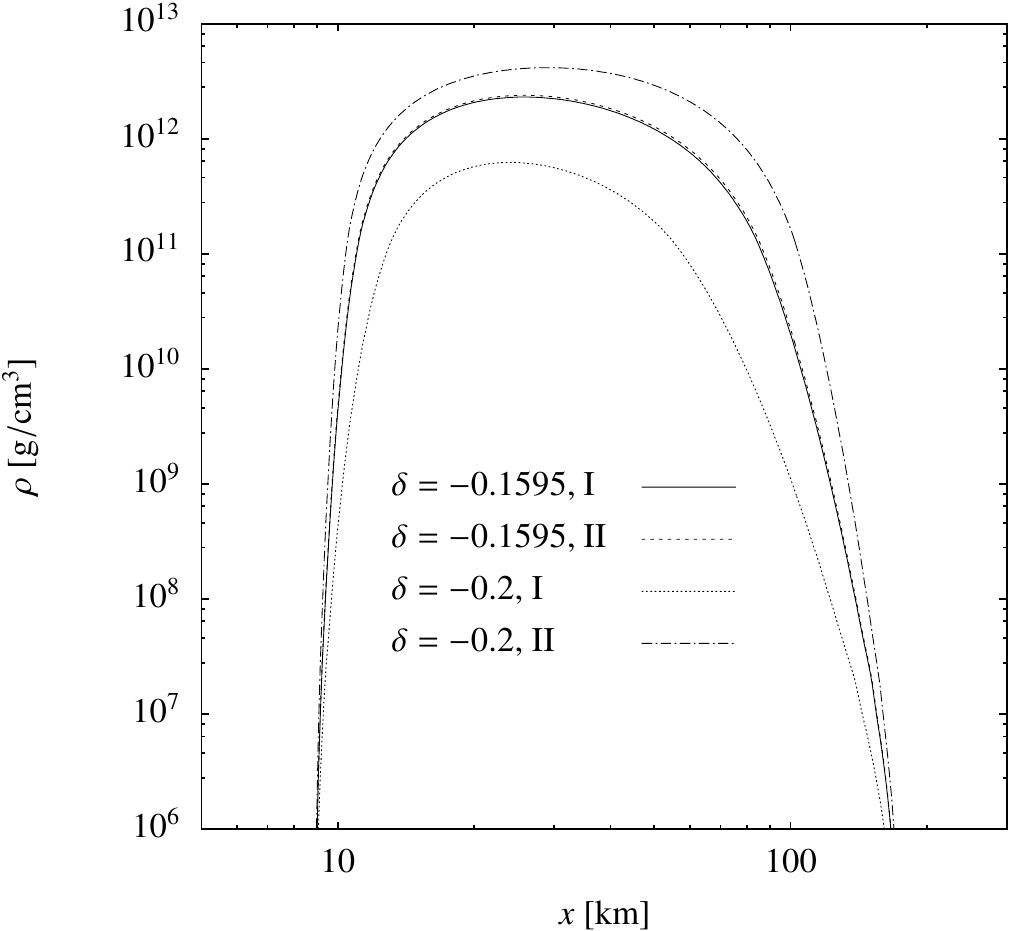}
\caption{Equatorial rest-mass density profiles for $s = 5 k$ near the bifurcation point $\delta^*_1 \approx -0.1595$. Here $r_1, r_2$, $a_\mathrm{rot}$, and $a$ are the same as in Fig.\ \ref{fig:Bifurcation_s}. The critical mass $M_\mathrm{disk}^*\in \left(0.40,0.43\right)M_\odot$. }
\label{fig:profil_s_5k}
\end{figure}  

 \begin{figure}[ht]
\includegraphics[width=1\columnwidth]{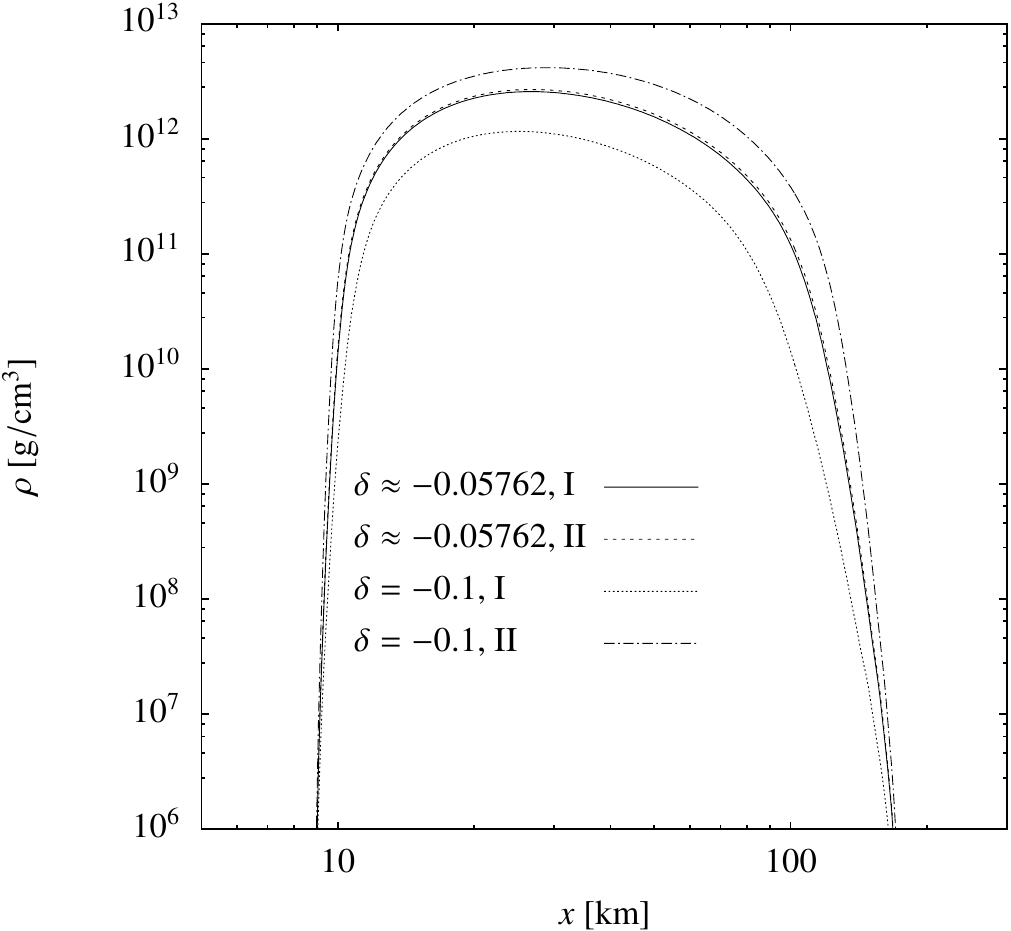}
\caption{Equatorial rest-mass density profiles for $s = 6 k$ near the bifurcation point $\delta^*_2 \approx -0.0576$. Here $r_1, r_2$, $a_\mathrm{rot}$, and $a$ are the same as in Fig.\ \ref{fig:Bifurcation_s}. The critical mass $M_\mathrm{disk}^*\in \left(0.86,0.93\right)M_\odot$.}
\label{fig:profil_s_6k}
\end{figure}

We failed to find any subcritical solutions---solutions seem to be  absent for values of the parameter $\delta $ that are larger than  the critical value $\delta^*_1$ ($ \delta^*_2$). It is interesting that in Fig.\ \ref{fig:Bifurcation_lin} the branches consisting of light solutions continue only up to a border value $\delta_\mathrm{b}\approx -1/3$. In contrast to that, the branch of heavy solutions approaches quite closely the points with $\delta =-1$.   

Figures \ref{fig:profil_s_5k}  and \ref{fig:profil_s_6k} display mass density profiles of disks that are close to  critical points $\delta_1^*$ or  $\delta_2^*$. Solid and dotted lines are denoted by the Roman numeral I; they depict solutions belonging to the light branch. Dashed and dashed-dotted lines, denoted by II, correspond to solutions that belong to the heavy branch. The two central curves (solid and broken lines) refer to the disk solutions that are very close to critical ones. The corresponding disk masses are given below. We have for configurations of Fig.\ \ref{fig:profil_s_5k}: on branch I---$0.40 M_\odot $ ($\delta = -0.1595 $) and $0.043 M_\odot $ ($\delta =-0.2$); on branch II---$1.29 M_\odot $ ($\delta =-0.2$) and $0.43 M_\odot $ ($\delta = -0.1595$). In the case of  configurations shown in Fig.\ \ref{fig:profil_s_6k}: ---$0.86M_\odot$ ($\delta \approx -0.0576$ )  and $0.22M_\odot$ ($\delta =-0.1$) on branch I; ---$2.1 M_\odot$ ($\delta =-0.1$) and $0.93 M_\odot$ ($\delta \approx-0.0576$ ) on branch II.

We failed to find a bifurcation diagram  in the important case when the critical solution corresponds to the Keplerian rotation ($\delta^* =-1/3$).   It probably does exist, but its critical point might exist for smaller $s$, presumably around $s = 4 k$;   we do not possess data defining the equation of state, in this case.

\section{Keplerian rotation laws and disks}
\label{Kepler}

\begin{figure}[ht]
\includegraphics[width=1\columnwidth]{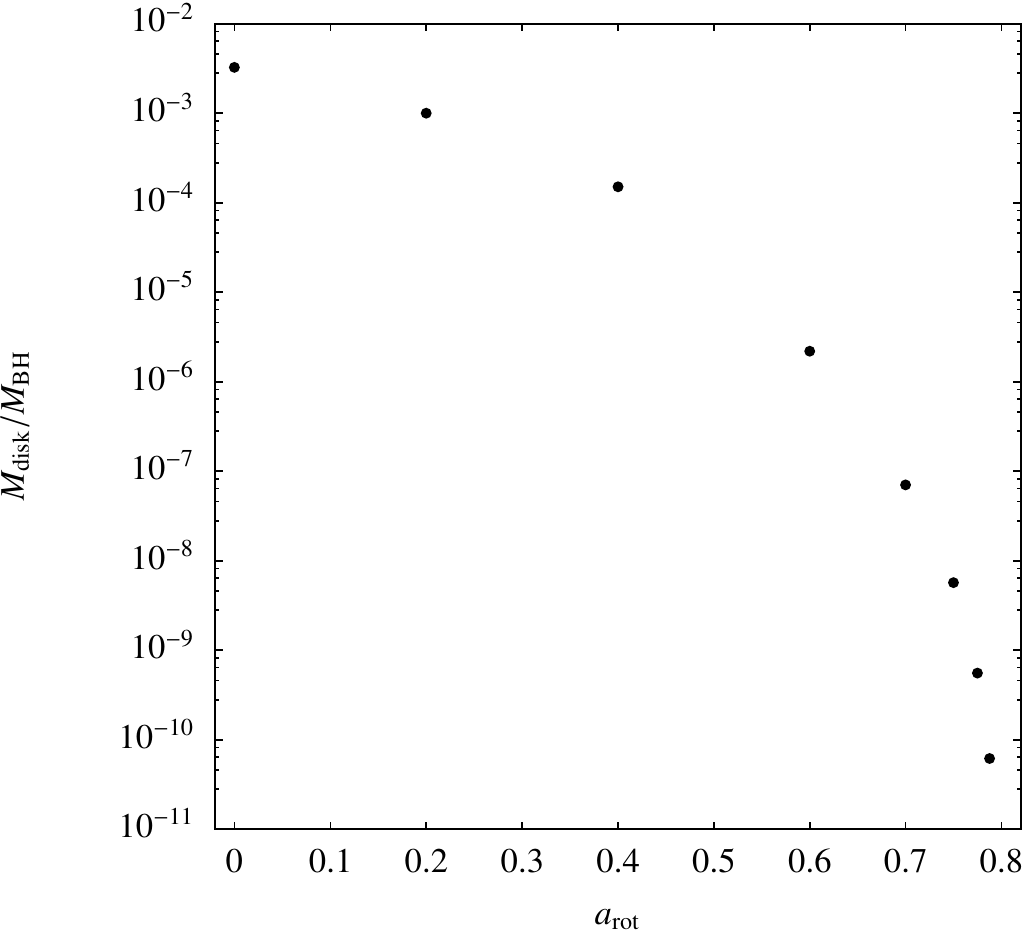}
\caption{The ratio $M_\mathrm{disk}/M_\mathrm{BH}$ vs.\ $a_\mathrm{rot}$ for light disks with the Keplerian rotation law (\ref{keplerian_rl}). Here $r_1=2$, $r_2=40$, $\delta =-1/3$, the mass parameter $m=1$, the spin parameter $a_\mathrm{rot} \in [0, 0.8)$, and $s=8k$. The black hole mass $M_\mathrm{BH}\approx 3M_\odot$.}
\label{fig:arotlaw_change}
\end{figure}

\begin{figure}[ht]
\includegraphics[width=1\columnwidth]{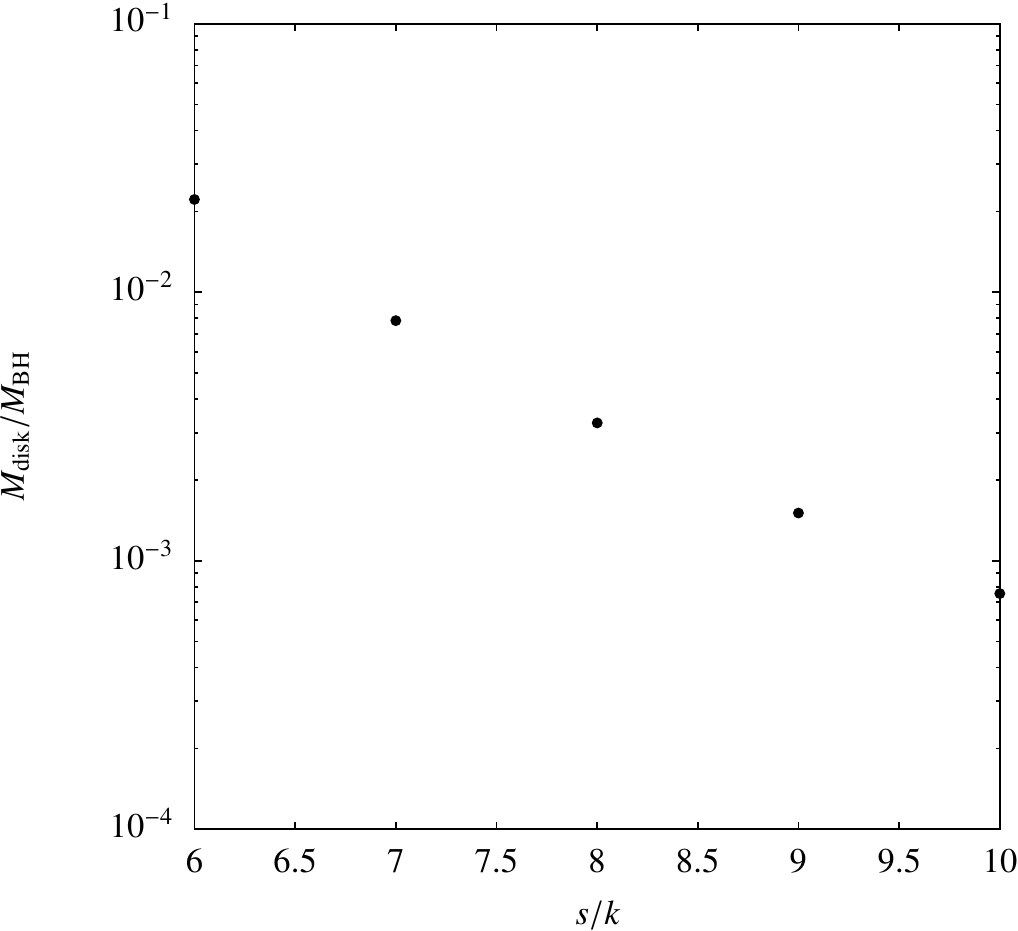}
\caption{The ratio $M_\mathrm{disk}/M_\mathrm{BH}$ vs.\ the specific entropy in the disk. The data are the same as in Fig.\ \ref{fig:arotlaw_change}, but now $a_\mathrm{rot}=0$, and the specific entropy changes: $s/k = 6,7,8,9,10$.}
\label{fig:arotlaw_change1}
\end{figure}

\begin{figure}[ht]
\includegraphics[width=1\columnwidth]{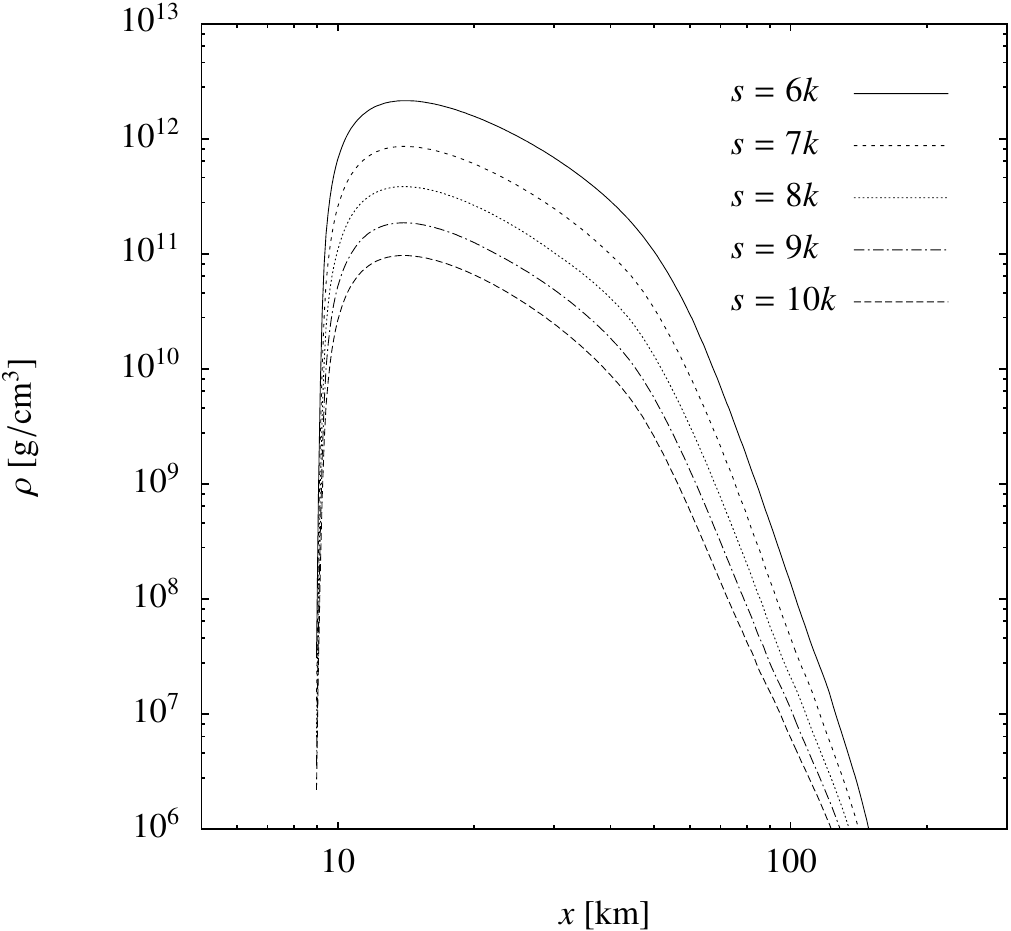}
\caption{Equatorial rest-mass density profiles for solutions shown in Fig.\ \ref{fig:arotlaw_change1}.}
\label{fig:arotlaw_change2}
\end{figure}

\begin{figure}[ht]
\includegraphics[width=1\columnwidth]{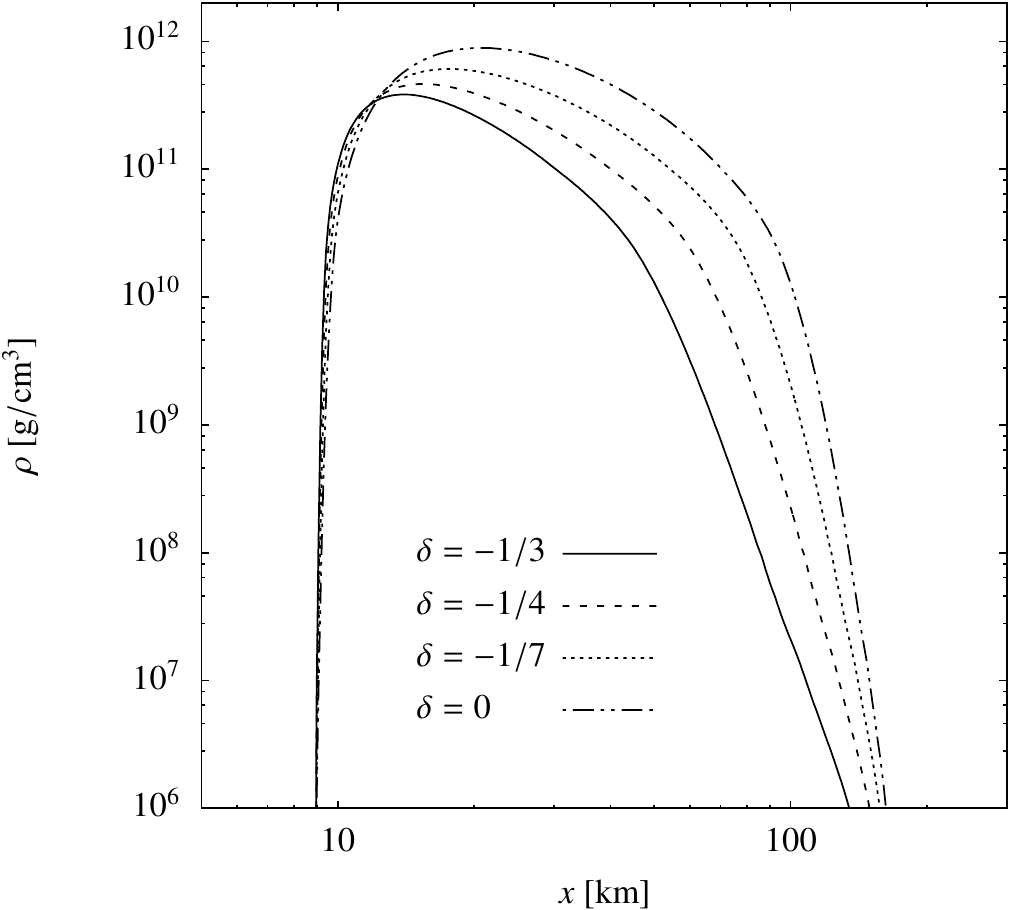}
\caption{Equatorial rest-mass density profiles of light disks obtained for $a_\mathrm{rot}=0$ and the rotation law (\ref{kmm}). The exponent $-\delta \in  \{0, 1/7, 1/4, 1/3\}$. The remaining parameters are the same as in Fig.\ \ref{fig:arotlaw_change}.}
\label{fig:arotlaw_0}
\end{figure}  

The Keplerian rotation  is common in rotating systems. It is the only allowed rotation in the case of massless disks of dust around a compact system, a black hole or a neutron star. It might well be so also for light gaseous disks, perhaps under certain restrictions concerning the equation of state. There exists a numerical indication that  the collapse of two neutron stars  can result in quasi-stationary tori rotating around a black hole with the ``almost'' Keplerian angular velocity   \cite{Rezzolla}. Such ``sufficiently compact'' stationary toroids have been absent---for the rotation law $j_\mathrm{Sh}$ with the exponent $-1/3$---in the analysis of \cite{sho_winter}. We shall investigate below a model with a special case of the rotation (\ref{keplerian_rl}), where the rotation parameter $a_\mathrm{rot}$ can differ from the spin parameter $a$. Our goal is to construct a numerical solution with a disk significantly less massive than the central black hole, but yet  sufficiently massive to be interesting astrophysically. 
  
In the first step we will study the impact of the parameter $a_\mathrm{rot}$ onto masses of disks for solutions belonging to branch I. In all cases shown below the coordinate size of disks is the same---the inner and outer coordinate radii are $r_1 = 2$, $r_2 = 40$ respectively. The spin parameter of the black hole is always $a = 0.8$.

Figure \ref{fig:arotlaw_change}, in which the specific entropy is relatively high, shows that masses depend quite significantly  on the value of $a_\mathrm{rot}$. The disk mass quickly decreases with the increase of $a_\mathrm{rot}$ and becomes a negligible entity (of the order of $10^{-11}$ of the mass of the black hole) when $a_\mathrm{rot}\approx 0.788$---a bit less than the spin of the  black hole, $a=0.8$. The heaviest disk corresponds to the parameter $a_\mathrm{rot}=0$, and its mass  $M_\mathrm{disk}=3.3\times 10^{-3}M_\mathrm{BH}$. That teaches us that we should choose $a_\mathrm{rot}=0$ and then try to maximize the mass of the disk by changing the specific entropy $s$. Results of this investigation are shown in Fig.\ \ref{fig:arotlaw_change1}, which displays masses of disks versus the specific entropy. In the case of $s = 5 k$  solutions have not been found, but they exist in the interval  $s\in [6 k, 10 k]$. The masses of tori decrease with the increase of the specific entropy. The largest disk mass is equal to about $0.02 M_\mathrm{BH}=0.06 M_\odot $, and it was obtained for $s = 6 k$. Relative disk masses decrease rapidly with the increase of $s$: we have $M_\mathrm{disk}/M_\mathrm{BH} = 7.8\times 10^{-3}$ for $s=7 k$, $M_\mathrm{disk}/M_\mathrm{BH} = 3.26\times 10^{-3}$ for $s = 8 k$, $M_\mathrm{disk}/M_\mathrm{BH} = 1.5\times 10^{-3}$ for $s = 9 k$, and $M_\mathrm{disk}/M_\mathrm{BH} = 7.5\times 10^{-4}$ for $s = 10 k$.

Figure \ref{fig:arotlaw_change2} shows mass density profiles of disks corresponding to solutions presented in Fig.\ \ref{fig:arotlaw_change1}. The maximal mass density of the heaviest disk, for which $s = 6 k$, reaches the value $2\times 10^{12}~ \mathrm{g/cm^3}$.
 
Finally, we shall show masses and density profiles of a Keplerian disk and a few non-Keplerian ones. Figure \ref{fig:arotlaw_0} compares various density profiles within light disks, for a family of rotation laws (\ref{kmm}) with the rotation parameter $a_\mathrm{rot}$ set to zero. The spin parameter of the black hole $a = 0.8$ is the same as in former examples. The black hole mass is $3M_\odot $. The general feature is that these profiles are shifted inwards with the decrease of $ \delta $. The values of the mass of the disk are given in Table \ref{tab:arotlaw_0}; they are decreasing with the decrease of $\delta$.

\begin{table}
\caption{The ratio $M_\mathrm{disk}/M_\mathrm{BH}$ for solutions presented in Fig.\ \ref{fig:arotlaw_0}.}
\begin{ruledtabular}
\begin{tabular}{c c c}
No.      & $\delta$ & $M_\mathrm{disk}/M_\mathrm{BH}$\\
\hline
1       & $-1/3$    & $3.26 \times 10^{-3}$\\
2       & $-1/4$    & $8.10 \times 10^{-3}$\\
3       & $-1/7$    & $2.17 \times 10^{-2}$\\
4       & $0$        & $5.78 \times 10^{-2}$\\
\end{tabular}
\end{ruledtabular}
\label{tab:arotlaw_0}
\end{table}

\section{Concluding remarks}
\label{conclusions}

We study stationary and axially symmetric black hole-disk systems. The fluid inside the disk satisfies the equation of state found by  Fujibayashi et al.\ \cite{sho_winter}---a combination of  the DD2 and Timmes-Swesty equations of state \cite{equation_of_state}. 

We choose two families of  rotation laws, that of  \cite{sho_winter} and of \cite{KM2020}. They are natural, for different reasons. The first is simple and yields well known  monomial Newtonian limits for the angular velocity. The second is formally more  complex, but it gives the right and exact answer in the case of a masless disk around a Kerr black hole \cite{kkmmop} (in such a case $\delta =-1/3$ and $\kappa =3$) and also gives monomial Newtonian limits for the angular velocity. 
We show, that the family of rotation laws of \cite{sho_winter}
is a subcase of that considered in \cite{KM2020}.

We present numerical evidence for the  existence of bifurcation.
There are two kinds of situations---the stronger case is when the parameter $\delta $ is treated as a bifurcation parameter. There is, however, a possibility that it is the  specific entropy that rules the bifurcation.  This emergence of bifurcation   is interesting mathematically, but there is also a potential physical application. Bifurcation often goes in pair, in nonlinear equations of physics, with the loss of stability of solutions. That would mean that there exists, in the context of gravitational coalescences of two compact  objects, a favoured (stable) branch of possible configurations consisting of a black hole and a toroid. A more technical reason why the investigation of bifurcation is important, is that the process of finding numerical solutions near a bifurcation point becomes rather subtle---it is easy to miss them.  Let us mention here a recent discovery of a   bifurcation in rotating polytropic disks \cite{dyba}, with yet another bifurcation parameter.  It is clear that the mathematics of rotating and self-gravitating matter within general relativity,
is not only complex but also rich in classic (nonlinear) phenomena. 

A new interesting feature of the family of rotation laws $j_\mathrm{KM}$ is that they allow for the existence of solutions with $\delta$ being equal to the ``Keplerian'' value $-1/3$, even for relatively compact configurations and light disks.    
There exists a regime within the DD2-Timmes-Swesty  equation of state of  a moderate  entropy, when  Keplerian disks can be moderately massive---up to 2\% (0.06 $M_\odot$) of the mass of the black hole.

\section*{Acknowledgments}
 We would like to thank Sho Fujibayashi and Masaru Shibata for making it possible to use their tabulated equation of state.

\end{document}